# Extreme magnetoresistance induced by Zeeman effect-driven electron-hole compensation and topological protection in MoSi$_2$

M. Matin,[1,2] R. Mondal,[1] N. Barman,[2] A. Thamizhavel,[1]* S. K. Dhar[1]

[1]Department of Condensed Matter Physics & Materials Science, Tata Institute of Fundamental Research, Mumbai-400005, India

[2]Department of Physics, Dinhata College, Dinhata-736135, West Bengal, India

E-mail: thamizh@tifr.res.in

## Abstract

The magnetoresistance is the magnetic field induced change of electrical resistivity of a material. Recent studies have revealed extremely large magnetoresistance in several non-magnetic semimetals, which has been explained on the basis of either electron-hole compensation or the Fermi surface topology, or the combination of both. Here, we present a single crystal study on MoSi$_2$, which exhibits extremely large magnetoresistance, approaching almost $10^7$ % at 2 K and 14 T magnetic field. It is found that the electron-hole compensation level in MoSi$_2$ evolves with magnetic field, which is resulted from strong Zeeman effect, and found beneficial in boosting the large non-saturating magnetoresistance. The non-trivial Berry phase in the de Haas-van Alphen oscillations and the moderate suppression of backward scattering of the charge carriers lend support for the topological nature of this semimetal. The ultra-large carrier mobility of the topologically protected charge carriers reinforces the magnetoresistance of MoSi$_2$ to an unprecedented large value.

## Introduction

Huge positive magnetoresistance (MR) of bismuth, a prototype elemental semimetal, has been known for a long time (*1*). Recently, the interest in MR has been renewed due to the observation of extremely large magnetoresistance (XMR) in WTe$_2$ which does not show saturation even in magnetic field as high as 60 T (*2*). The XMR in WTe$_2$ was followed by similar observations in several other semimetals. For example, an XMR~$4\times10^6$ % at 2 K and 14 T has been reported in PtBi$_2$ (*3*). The other semimetals with comparable XMR are NbP, LaSb, NbAs, TaP, TaAs, MoTe$_2$, and MoP$_2$ (*4-10*). While the members of the TrX (*Tr* = Nb, Ta; X = P, As) family have been experimentally suggested to be Weyl semimetals (*4, 11-13*), WTe$_2$, MoTe$_2$, MoP$_2$, and PtBi$_2$ belong to a special class of type-II Weyl semimetals (*3, 9, 10, 14-16*), in which the Fermi arc is slightly tilted with respect to the Fermi level. As a result of their exotic electronic structure, the type-II Weyl semimetals host a combination of ultra-large carrier mobility and electron-hole compensation, and thus present a new platform to realize huge MR.

In late 1986, MoSi$_2$ was reported to exhibit strong MR effect (*17*), almost comparable to those observed in type-II Weyl semimetals. The resonant compensation between electrons and holes was predicted to be the underlying mechanism behind the origin of XMR in MoSi$_2$, though this consensus is far from being proved experimentally. Despite significant new understanding of the MR phenomenon developed over the last few years, no further study had been attempted to check whether the notion of topological protection can be extended to MoSi$_2$, which seems meaningful, at least looking at the large MR exhibited by this semimetal. Our magnetotransport study on an exceptionally good quality single crystal of MoSi$_2$ reveals that there exists an imbalance between the electron and hole densities, which is in stern contrast with the previous consensus. However, the Zeeman effect causes a magnetic field induced-modulation of the Fermi surface, and drives the system towards resonant compensation condition. The non-trivial Berry phase shift in the de Haas-van Alphen oscillations and the moderate suppression of backward scatterings provide experimental evidences in favor of the topologically non-trivial state in MoSi$_2$. The combination of the magnetic field-driven resonant compensation and the topological protection leads to an unprecedented large non-saturating MR in MoSi$_2$.

## Results and discussion

Figs. 1A and B show the temperature dependence of electrical resistivity $\rho_{xx}$ of MoSi$_2$ measured from 2 to 300 K in few representative transverse magnetic fields ranging from zero to 14 T. Zero-field resistivity exhibits metallic behavior, with $\rho_{xx}$ decreasing steeply as temperature is lowered from 100 to 15 K, the drop being steeper for $I\|a$. The steep decrement in $\rho_{xx}$ implies a strong enhancement in the transport lifetime of the charge carriers. For $I\|a$, this enhancement results in an extremely small residual resistivity $\rho_0$=7.1 nΩ-cm and a large residual resistivity ratio (RRR) ~3926. The present RRR value is more than twice the best RRR value reported earlier for this compound (*17*), attesting a better quality of our MoSi$_2$ single crystal. For $I\|c$, $\rho_0$ is 48.8 nΩ-cm and the corresponding RRR is 336, implying a relatively smaller transport lifetime of the charge carriers along this crystallographic axis. At the lowest temperature, the anisotropy ratio $\rho_c/\rho_a$ is ~7. The anisotropic behavior of resistivity is also observed in the high-temperature regime, but with $\rho_c/\rho_a <$ 1.

The application of a transverse magnetic field leads to a remarkable enhancement in $\rho_{xx}$, particularly at low temperatures, and the MR is observed to be larger when $I\|a$. For applied magnetic field larger than a critical magnetic field $B^*$~0.3 T for $I\|c$ and ~0.15 T for $I\|a$, $\rho_{xx}(T)$ exhibits a metal to insulator-like crossover upon lowering the temperature. At magnetic fields above 9 T, a resistivity plateau is eventually formed after an inflection at around 15 K. For magnetic field $\leq$ 9 T, the plateau region is replaced by a re-entrant metallic state, as shown in Fig. 1C for few selected magnetic fields. While the metal-to-insulator-like crossover is considered to be a unique characteristic of semimetals possessing high carrier mobility (*4, 5, 9*), the observation of the re-entrant metallic state has been reported only in the elemental semimetals like graphite, bismuth, and antimony (*18, 19*). MoSi$_2$ exhibits extremely large transverse MR [=($\rho_{xx}(T,B)/\rho_{xx}(T,0)$ –1)×100 %] at low temperatures. For $I\|a$, the MR reaches a value as large as 9.1×10$^6$ % at 2 K and 14 T magnetic field. This value is much larger than the MR reported in the recently discovered XMR materials like PtBi$_2$, WTe$_2$ and LaSb (*2, 5, 8*), and comparable to the MR in the prototype elemental semimetal Bi (*20*). The present XMR value exceeds the previously reported value on MoSi$_2$ by almost one order of magnitude (*17*), which further reflects the better quality of our single crystal. The transverse MR is relatively lower when measured with the current along the *c*-axis, and assumes a value of 3.8×10$^5$ % at 2 K and 14 T magnetic field. This value is still comparable with the MR observed in NbP and TaAs$_2$ (*4, 21*).

The magnetic field dependence of transverse MR measured at several selected temperatures between 2 and 100 K with current parallel to the *a*-axis is shown in Fig. 1D. The MR exhibits non-saturating behavior up to the highest applied magnetic field of 14 T. It decreases slowly with increasing temperature from 2 to 15 K, but drops rapidly at higher temperatures. At 100 K, MR reduces to a value of only ~539 % at the highest magnetic field. The MR at 2 K shows nearly quadratic field dependence, i.e., MR∝$B^{1.92}$. Such behavior is reminiscent of compensated semimetals. Qualitatively similar behavior of MR is observed for *I*||*c*, and is shown in the supplementary Fig. S1.

The magnetic field dependence of magnetization (*M*) of $MoSi_2$ displays robust signature of the de Haas-van Alphen (dHvA) oscillations superimposed on a diamagnetic background. The dHvA oscillation signal is extracted by subtracting the non-oscillatory background from the experimental *M(B)* data. As shown in Fig. 2A, the dHvA oscillations measured at 2 K and with magnetic field applied along the *c*-axis are traceable down to remarkably low magnetic field ~ 1.7 T. The oscillation amplitude diminishes gradually upon increasing temperature, and becomes hardly detectable above 15 K. As shown in Fig. 2B, the dHvA oscillations at low temperature (say 2 and 3 K) exhibit strong Zeeman effect, which manifests itself as splitting in the oscillation peaks. At higher temperatures, the split peaks merge into a single peak, consistent with the Zeeman effect, in which the thermal broadening of the Landau levels smears out the Zeeman splitting. The Zeeman splitting at 2 K is discernible down to almost 4 T, which is much smaller than the threshold field value B > 17 T for peak splitting in topological semimetallic candidate $Cd_3As_2$ with the highest carrier quantum mobility reported so far (*22-24*). Such strong Zeeman splitting can only be realized in topological semimetals. Fig. 2C shows the frequency spectra inferred from the dHvA oscillations, revealing two principal frequencies: $F_\alpha$ = 725 T and $F_\beta$ = 825 T, which are in close agreement with the previous experiment (*25*). The temperature dependence of the dHvA oscillation amplitude shown in Fig. 2D, is fitted using the thermal damping factor of the Lifshitz-Kosevich formula (*26*): $R_T = X/\sinh(X)$, X = $\lambda T m^*/B$, where $\lambda = 2\pi^2 k_B m_e/e\hbar$ is a constant term containing the Boltzmann constant $k_B$, electron charge *e*, and reduced Planck's constant $\hbar$. Here, $m^*$ is the cyclotron (effective) mass in the unit of free electron mass $m_e$. The fitting parameter provides a determination of $m^*$. From the magnetic field-induced damping of the oscillation amplitude, $R_B = \exp(-\lambda T_D m^*/B)$, the Dingle temperature $T_D$ at 2 K is estimated; the corresponding fits are shown in the inset of Fig. 2D. The other parameters of interest are obtained as follows: the quantum lifetime $\tau_Q = \hbar/2\pi k_B T_D$ and the quantum mobility $\mu_Q = e\tau_Q/m^*$. The estimated parameters for the two Fermi pockets are summarized in Table-1.

The electronic band structure shown in Fig. 3A indicates that there are two bands crossing the Fermi level, consistent with the dHvA oscillations result. The hole-type and the electron-type bands are centered at the high symmetric *Γ*- and *Z*-point, respectively. As shown in Fig. 3B the hole-type band has an almost cylindrical-shaped Fermi surface while the electron-type band has Fermi surface with a shape of a four-cornered rosette. The small indirect overlap between electron-type and hole-type bands reinforces the semimetallic character of $MoSi_2$.

Having established plausible evidence for the presence of two bands in $MoSi_2$, we now use the classical theory of two-band transport for analyzing the magnetransport data. In the two-band model, the transverse MR and Hall resistivity $\rho_{xy}$ are given by (*27*):

$$\frac{\Delta \rho}{\rho} = \frac{(n\mu_e + p\mu_h)^2 + (n\mu_e + p\mu_h)(p\mu_e + n\mu_h)\mu_e\mu_h B^2}{(n\mu_e + p\mu_h)^2 + (p-n)^2 \mu_e^2 \mu_h^2 B^2} - 1 \quad (1)$$

$$\rho_{xy} = \frac{1}{e}\left[\frac{(p\mu_h^2 - n\mu_e^2)B + (p-n)\mu_e^2\mu_h^2 B^3}{(n\mu_e + p\mu_h)^2 + (p-n)^2 \mu_e^2 \mu_h^2 B^2}\right] \quad (2)$$

where, $n$ ($p$) is the electron (hole) density and $\mu_e$ ($\mu_h$) the corresponding mobility. Conventionally, the MR and Hall resistivity data are simultaneously fitted to the above equations for obtaining precise values of the carrier density and mobility. The presence of four model parameters ($n$, $p$, $\mu_e$ and $\mu_h$), appearing as multiplicative pairs, however, enhances the possibility of ambiguous determination of the carrier density and mobility. Our practical experience suggests that any arbitrary guess value of one of the model parameters in the physically acceptable range reproduces the experimental data by adjusting the values of the other model parameters, with a variation large enough to deter us from using eqn. (1) as such to analyze the MR data. Considering the non-saturating MR of $MoSi_2$ with nearly quadratic field dependence, it is reasonable to assume that the compound is close to the resonant compensation condition. In such a case, tuning of the difference between $\mu_e$ and $\mu_h$ without altering the value of their product $\mu_e\mu_h$, does not appreciably change the calculated result, which allows one to replace $\mu_e$ and $\mu_h$ by a single parameter $\mu$ with the only constraint that $\mu \sim (\mu_e\mu_h)^{1/2}$. Therefore, in the limit $n \sim p$, eqn. (1) reduces to

$$\frac{\Delta \rho}{\rho} = \frac{(1 + p/n)^2(1 + \mu^2 B^2)}{(1 + p/n)^2 + (p/n - 1)^2 \mu^2 B^2} \quad (3)$$

Eqn. (3) involves only $\mu$ and the compensation level $r = p/n$ as model parameters. Here, we emphasize that it is the compensation level $r$ rather than the actual values of the carrier densities, which plays the fundamental role in determining the MR in compensated semimetals. Another important point to be noted here is that the functional form of eqn. (3) remains unaltered upon replacing $p/n$ by its reciprocal. Thus, the two distinctly different values of $r$, namely $p/n$ and $n/p$ will produce the same fitting curve for a fixed value of $\mu$. Fig. 4A shows that eqn. (3) nicely reproduces the MR data for $I\|c$. The best fit to the 2 K-MR data is achieved with $r = 0.988$ and the geometrically average mobility $\mu \sim (\mu_e\mu_h)^{1/2} = 6.3 \times 10^4$ cm$^2$/V-s. As argued above, the same 2 K-fit could also be reproduced with $r = 1/0.988 = 1.012$. However, the negative sign of the Hall resistivity for $I\|c$ and $B\|a$ shown in Fig. 4B evidences a charge transport mechanism with majority electron carriers along the $c$-axis, imposing a tight constraint on $p/n$, and allows one to assign the value 0.988 to $p/n$ without ambiguity.

The Hall resistivity $\rho_{xy}$ for $I\|a$ and $B\|c$ shown in Fig. 4B is also negative at low temperature, with similar implication as for $I\|c$. According to two-band model eqn. 2, $\rho_{xy}(B)$ should exhibit linear behavior in case of perfect electron-hole compensation. If some imbalance exists between the electron and hole densities, $\rho_{xy}(B)$ accumulates a sub-linear character with negative curvature, which is apparently at odds with what is observed experimentally for $I\|a$ and $B\|c$. The peculiar behavior of $\rho_{xy}(B)$ with a distinct positive curvature in high-field regime cannot be reconciled by tuning the carrier mobilities and densities, albeit an assumption of gradually increasing contribution of the holes in the charge transport mechanism can explain the observed behavior. Similar behavior of $p/n$ has been inferred for $WTe_2$ (28, 29), where the Fermi surface is predicted to exhibit a strong magnetic-field-dependent behavior because of the Zeeman effect. The Subnikov-de Haas (SdH) oscillations are also probed through resistivity measurements (see Fig. *S*2 and the relevant text in the supplementary materials), which also demonstrate strong Zeeman splitting of the SdH oscillations peaks for $I\|a$ and $B\|c$. However, such peak splitting is completely absent in the SdH oscillations for $I\|c$ and $B\|a$. It is worth noting that $\rho_{xy}(B)$ for $I\|c$ and $B\|a$ exhibits conventional behavior of two-carrier system without featuring any positive curvature. These observations conclusively establish a correlation between the Zeeman effect and the magnetic field-induced evolution of $p/n$.

The description of the magnetotransport data within the two-band model is admissible as long as the model parameters are independent of magnetic field, which seems not to be a good approximation for $I\|a$. This compels us to perform the fitting of the experimental data within a small magnetic field range where the variation of $p/n$ is rather marginal, minimizing the error in the derived values of the fitting parameters. The fitting of the MR data at 2 K in the magnetic field range up to 2 T is achieved with $r = 0.983$ and $\mu = 3.4 \times 10^5$ cm$^2$/V-s (Fig. 4C). The fitted curve when extrapolated to higher magnetic fields exhibits strong tendency of saturation, furnishing a maximum MR at 14 T of only $1.5 \times 10^6$ %, which is significantly lower than the experimentally observed value. As the magnetic field is increased above 2 T, it causes a gradual increment in $p/n$ from its value of 0.983 prevailing in low-field regime, driving the system towards the resonant compensation condition. In that sense, the magnetic field plays a crucial role in boosting the MR. Once the resonant compensation condition is reached, further increase of magnetic field increases the $p/n$ further, taking the system away from the resonant compensation condition. This well explains the mild saturation of the MR observed in high-field regime above 5 T in the *log-log* plot of MR($B$) data shown in Fig. 4D.

Another important feature of $\rho_{xy}(B)$ for $I\|a$ and $B\|c$ is its exhibition of sign reversal from positive at low-field to negative at high-field regime (Fig. 4E). The zero-crossing field $B_c$ at 2 K is ~228 mT, and it increases progressively with increasing temperature. According to the two-band model eqn. (2), this zero-crossing occurs at $B_c = \pm[(r\mu_h^2 - \mu_e^2)/\mu_h^2\mu_e^2(1-r)]^{1/2}$, which imposes a quantitative constraint on the model parameters. This constraint together with the values of $r$ and $\mu_e\mu_h$ obtained from the MR data provides a rough estimation of $\mu_e$ and $\mu_h$. For $I\|a$, the estimated parameters at 2 K are: $\mu_e = 2.5 \times 10^5$ cm$^2$/V-s and $\mu_h = 4.6 \times 10^5$ cm$^2$/V-s. In the final stage of refinement of the model parameters, the obtained values are taken as initial guess values and allowed to vary slightly to reproduce the $\rho_{xy}(B)$ data at 2 K in the magnetic field range 0-2 T (Fig. 4F). The fitting provides the final set of values of the model parameters as: $n = 1.75 \times 10^{21}$ cm$^{-3}$, $p = 1.67 \times 10^{21}$ cm$^{-3}$, $\mu_e = 2.7 \times 10^5$ cm$^2$/V-s and $\mu_h = 4.5 \times 10^5$ cm$^2$/V-s. Our results indicate that MoSi$_2$ cannot be considered as a perfectly compensated semimetal as predicted earlier (*17*, *25*). In low-field limit, electrons and holes are rather largely uncompensated, but magnetic field improves the compensation level and functions to boost the non-saturating XMR in MoSi$_2$ in a non-trivial way.

While the ultra-large carrier mobility and the accompanied XMR in MoSi$_2$ may be an indication of high purity of the sample, another intriguing possibility for this is the topological protection that severely suppresses the backward scatterings. The experimental proof of the suppression comes from the ratio $R_\tau$ of the transport lifetime $\tau_{tr}$ to the quantum lifetime $\tau_Q$. Under severe suppression, $R_\tau \gg 1$ as is the case for Dirac semimetallic candidate Cd$_3$As$_2$ ($R_\tau \sim 10^4$); see ref. (*30*). Using $m^* \sim 0.3 m_e$ and $\mu_h = 4.5 \times 10^5$ cm$^2$/V-s, the $\tau_{tr}$ (=$m^*\mu/e$) is estimated to be $7.2 \times 10^{-11}$ s for $I\|a$, giving $R_\tau \sim 165$. This ratio is smaller compared to those in MoP$_2$ ($R_\tau \sim 5 \times 10^3$) and NbAs ($R_\tau \sim 10^3$) (*10*, *11*), but comparable or even larger than the values reported in other topological semimetals like PtBi$_2$, and TaAs (*3*, *31*).

We explore further the possible topological character of MoSi$_2$ from an analysis of the Berry phase. A Berry phase shift of $\pi$ accumulated in the cyclotron motion of the Dirac/Weyl fermions constitutes another fundamental property of the topological semimetals (*32*). The presence of more than one frequency branch in the quantum oscillations in MoSi$_2$ produces large uncertainty in indexing the Landau levels, making the determination of Berry phase shift from Landau level fan diagram completely impractical. Thus, we choose to fit the dHvA oscillations to the Lifshitz-Kosevich formula for multiple frequencies given as (*26*, *33*):

$$\Delta M = \sum_i a_i B^{\frac{1}{2}} R_T^i R_B^i R_S \sin[2\pi(F_i/B - \gamma_i)] \quad (4)$$

where, the terms $R_T$ and $R_B$ describe respectively the temperature and magnetic field damping of the oscillation amplitude, respectively; and $R_S = \cos(\pi m^* g/2m_e)$ is the spin reduction factor due to the Zeeman splitting, which when constant, has no effect in peak splitting in the simple Lifshitz-Kosevich formula presented by eqn. (4) (*g* is the Landè *g*-factor). The phase factor $\gamma$ is related to the Berry phase factor $f_B$ as $\gamma = 1/2 - f_B + \delta$, where $\delta$ is an additional phase factor determined by the dimensionality, taking the values of 0 and $\pm 1/8$ for two and three-dimensional Fermi surface, respectively (plus and minus signs are respectively for hole and electron) (*34*). The Berry phase factor $f_B$ assumes the values of 0 or 1/2 depending on whether the material is topologically trivial or not. Starting from the previously estimated values of $F$, $m^*$ and $T_D$ for the $\alpha$- and $\beta$-frequency branches, we perform a fully quantitative fitting of the quantum oscillations of magnetization at 8 K with eqn. (4). For the fitting, we restricted ourselves to only high temperature and in low magnetic field regime, where the experimental data are less likely to be affected by the Zeeman effect. As evident from Fig. 5A, the experimental oscillation pattern can be well reproduced by the Lifshitz-Kosevich formula in narrow magnetic field range between 6 and 7 T with the values of the parameters $F_i$, $m_i^*$ and $T_D^i$ that are close to their initial guess values (variation being less than 7 %). The fitting gives the values of $\gamma_\alpha = -0.05$ and $\gamma_\beta = 0.23$. The $\alpha$-band represents an electron-type band as indicated by the negative sign of $\gamma_\alpha$. The electron-pocket shown in Fig. 3B has a quasi-two-dimensional nature, limiting the value of $\delta_\alpha$ in the range of $-1/8 < \delta_\alpha < 0$. Obviously, the fitted value of $\gamma_\alpha$ deviates largely from the value of $1/2 < \gamma_\alpha < 0.375$ expected for topologically trivial case with $f_B = 0$, and lies in the range of $-1/8 < \gamma_\alpha < 0$ expected for the topologically non-trivial electrons with $f_B = 1/2$. For the $\beta$-band, which represents a hole-type band, $\gamma_\beta$ shows strong deviation from the expected value of $(1/2 - 1/8) = 0.625$ for topologically trivial case. However, it also deviates slightly from the three-dimensional limiting value of $\gamma_\beta = \delta_\alpha = +1/8$ expected for the topologically non-trivial holes, and corresponds to a Berry phase shift of $0.8\pi$. This marginal deviation is likely due to cylindrical Fermi surface of the hole-pocket instead of spherical one with perfect $\pi$ Berry phase. Fig. 5A also shows that the deviation between the experimental data and theory becomes apparent as we move away from the aforementioned fitting field range. Since the carrier density is related to the principal frequency, its variation resulting from the change in magnetic field contributes to such deviation.

Lastly, we would like to point out another feature of our results that lends further support to the topological nature of MoSi$_2$. As shown in Fig. 5B, the low-temperature magnetization measured with magnetic field applied along the *c*-axis, exhibits the diamagnetic behavior throughout the measured field range except in the vicinity of zero field where a clear paramagnetic contribution is observed. The corresponding differential susceptibility $\chi$ ($=dM/dB$) shown in the inset of Fig. 5B demonstrates paramagnetic singularity near zero field, varying linearly with magnetic field. Similar low-field paramagnetic singularity of $\chi$ with a linear magnetic field dependence has been reported in three-dimensional topological insulators such as $Bi_2Se_3$, $Bi_2Te_3$, $Sb_2Te_3$, LaBi, $Bi_{1.5}Sb_{0.5}Te_{1.7}Se_{1.3}$, and $ZrTe_5$ (*35-37*). These materials host topologically protected surface state. Owing to the particular helicity of the surface state Dirac cones, the upper Dirac cone possesses a left-handed spin texture whereas the lower one possesses a right-handed spin texture. The spins of the electrons at the Dirac point do not have any preferable orientation and are free to align along the magnetic field direction as long as the Dirac cone is not gapped. These freely oriented spins are predicted to give rise to the paramagnetic singularity of $\chi(B)$ in low-field regime (*35*). Interestingly, such spin texture may also lead to the non-trivial Berry phase and suppression of backward scattering, and these concomitant phenomena are indeed observed in $MoSi_2$. The paramagnetic behavior of $\chi(B)$ in $MoSi_2$, however, remains unaltered upon changing the surface area of the sample, discarding surface state origin of the effect. It should be pointed out that similar spin texture may also exist in the bulk state of material with strong spin-orbit coupling, such as in BiTeI and $WTe_2$ (*34*, *38*). We believe that, similar to $WTe_2$, the spin-orbit coupling and the related spin texture may be accounted for the topological behavior of $MoSi_2$.

## Summary

Our magnetotransport results in $MoSi_2$ reveal that there exists an imbalance between the electron and hole densities, which is sufficient to cause saturation of magnetoresistance even in moderate magnetic field. Nevertheless, $MoSi_2$ exhibits extremely large magnetoresistance of the order of $10^7$ % at 2 K and in 14 T magnetic field without appreciable saturation. An intriguing aspect of the $MoSi_2$ is its exhibition of strong Zeeman effect which has a general consequence of magnetic field-induced reconstruction of the Fermi surface. Consequently, in $MoSi_2$, application of magnetic field improves the compensation level, driving the system towards the resonant compensation condition. This unique mechanism together with the ultra-large carrier mobility makes $MoSi_2$ a material with extremely large non-saturating magnetoresistance. The non-trivial Berry phase shift in de Haas-van Alphen oscillations and the moderate suppression of the backward scatterings provides evidences for the topologically non-trivial state in $MoSi_2$, and thus unfolds another interesting feature of this semimetal. The low-field paramagnetic singularity of magnetic susceptibility suggests that spin texture possibly exists in the bulk state of $MoSi_2$, which may lend a topological character to this semimetal. However, our electronic band structure calculation does indicate that the spin-orbit interaction − a key ingredient for realizing such spin texture, is not of much importance in case of $MoSi_2$. A direct probe like angle and/or spin resolved photoemission experiment is desirable for unveiling the origin of the topological character of $MoSi_2$.

## Materials and methods

Single crystal of $MoSi_2$ was grown by the Czochralski method using elemental molybdenum and silicon with purity levels of 99.99 % and 99.999 %, respectively. The phase purity of the as grown single crystal was checked by the powder X-ray diffraction experiment performed using CuK*α* radiation on a PANalytical X-ray diffractometer. A standard four-probe method was employed to measure the electrical resistivity in the temperature range 2–300 K and in transverse magnetic fields up to 14 T. The Hall resistivity in presence of fields up to 14 T was measured in several selected temperatures between 2 and 100 K. For the Hall resistivity measurements, we adopt the five-terminal configuration because it minimizes the offset voltage arising from the possible misalignment of the voltage leads. These electrical transport measurements were performed in a Quantum Design physical property measurement system. The magnetization measurements are performed in a vibrating sample magnetometer from Quantum Design, which is equipped with a superconducting magnet that can produce a magnetic field up to 14 T in liquid helium temperature. The electronic structure calculations were performed using the experimental lattice parameters (inferred from the X-ray diffraction pattern) within the density functional theory framework, as implemented in the Quantum espresso code (*39*). The generalized gradient approximation of Perdew *et al.* (*40*) was used for the exchange-correlation potential. The electronic wave-functions were expanded in terms of plane-wave basis sets, and pseudo-potential approach has been adopted for the calculations. The Fermi surface topology was plotted with the program Xcrysden.

## Acknowledgments

**General**: We thank Arvind Maurya and Dai Aoki for insightful conversations. We are also thakful to Sindhunil Barman Roy for critical reading of the manuscript. **Author contributions:** All the authors except N. B. have equal contribution in analyzing the magnetotranport and dHvA data presented in this study. The author N. B. has performed the electronic band structure calculations.

## Figures and tables

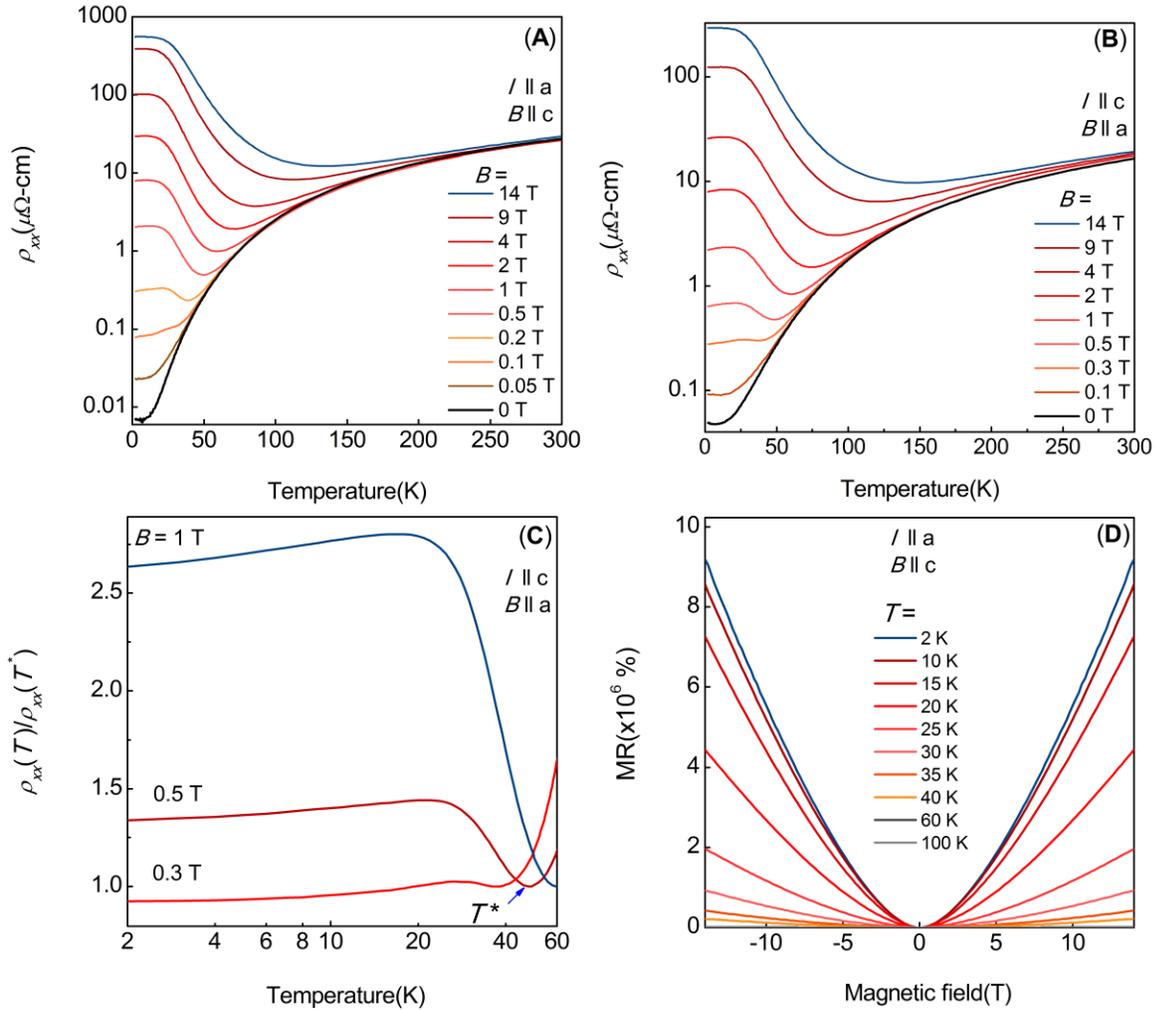

**Fig. 1. Magnetotransport properties of MoSi$_2$.** (**A**,**B**) Temperature dependence of the electrical resistivity $\rho_{xx}$ measured with current along the *c*- and *a*-axis under few representative transverse magnetic fields up to 14 T. (**C**) Temperature dependence of normalized resistivity for $I \parallel c$ under 0.3, 0.5 and 1 T magnetic fields, demonstrating the crossover to a re-entrant metallic state. (**D**) Magnetic field dependence of transverse MR for $I \parallel a$ at few selected constant temperatures between 2 and 100 K.

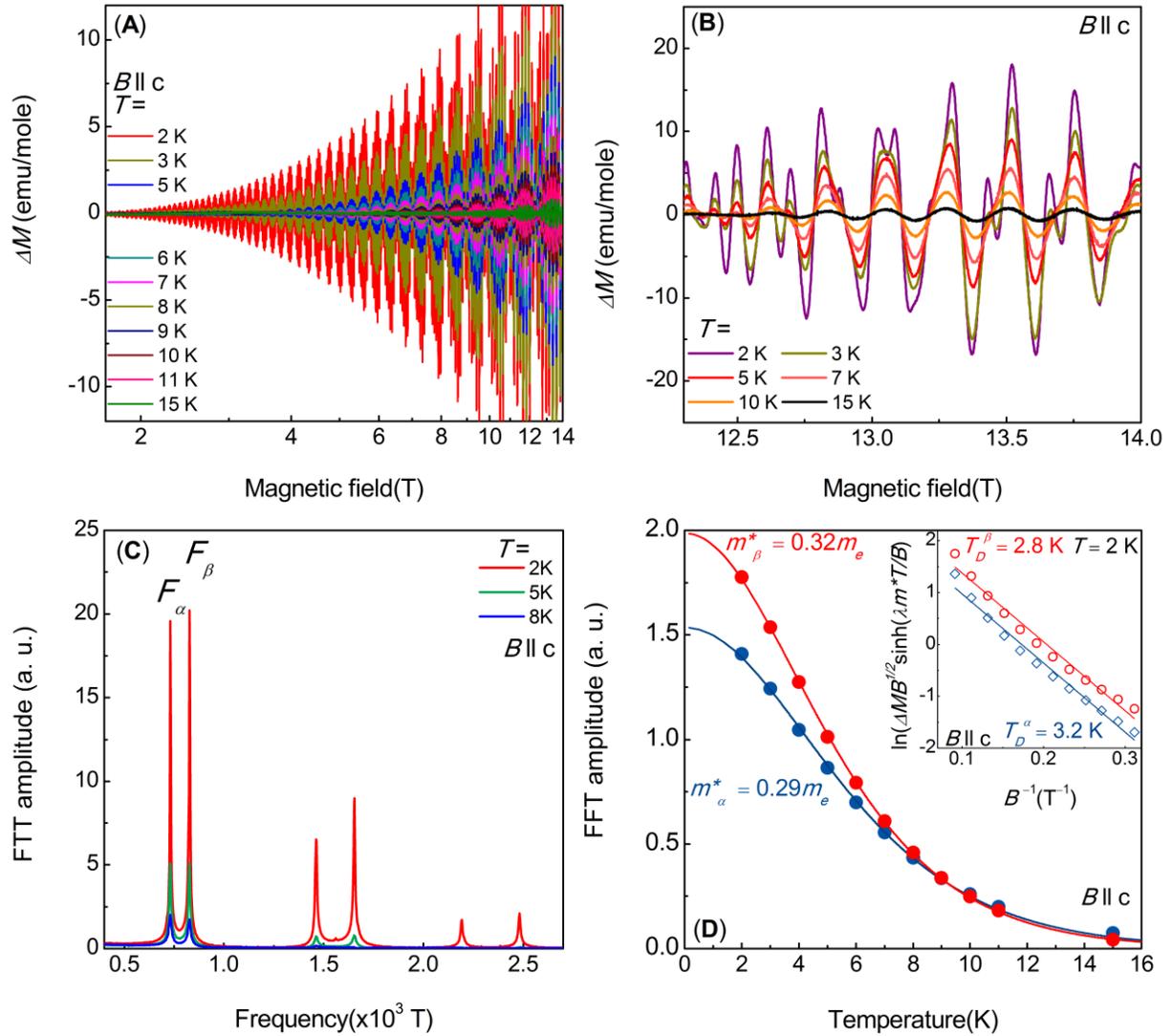

**Fig. 2. de Haas-van Alphen (dHvA) quantum oscillations in MoSi$_2$**. (**A**,**B**) The dHvA oscillations of magnetization at few selected constant temperatures between 2 and 15 K for magnetic field applied along the *c*-axis. The oscillation peaks at low temperature and in high magnetic field demonstrate the Zeeman splitting. (**C**) The frequency spectrum corresponding to the dHvA oscillations, revealing the presence of the principle frequencies $F_\alpha$ = 725 T and $F_\beta$ = 825 T along with their second, third and fourth harmonics. (**D**) Temperature dependence of the FTT amplitude at frequencies $F_\alpha$ and $F_\beta$ (solid symbols). The solid lines are the fitted curves using the thermal damping factor of the Lifshitz-Kosevich formula (see the relevant text). **Inset:** The Dingle plots of the dHvA oscillations associated with the frequencies $F_\alpha$ and $F_\beta$.

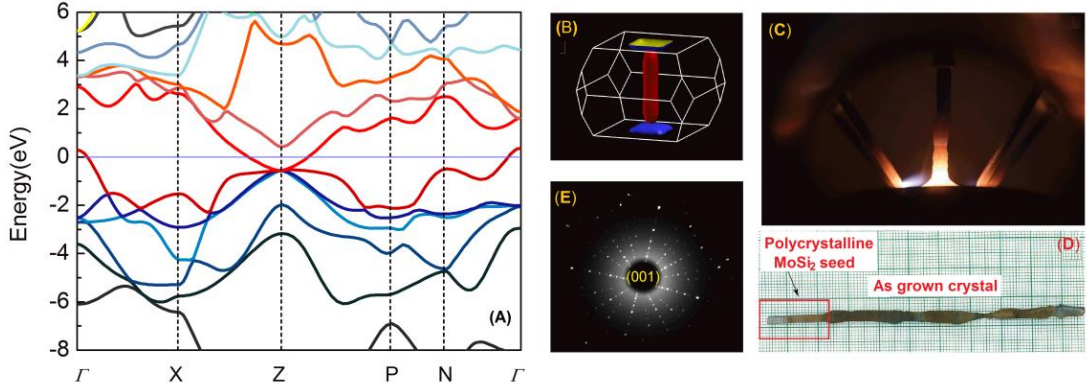

**Fig. 3. Electronic band structure of MoSi$_2$ and a brief overview of the single crystal growth**. (**A**) Electronic band structure of MoSi$_2$ with one electron-type band that crosses the Fermi level (indicated by the zero energy line) near the $\Gamma$-point and one hole-type band that crosses the Fermi level near the Z-point. (**B**) The Fermi surfaces of MoSi$_2$ in the first Brillouin zone with a central hole pocket (cylindrical) and the peripheral electron pocket (having shape of a four-cornered rosette). (**C,D**) Snap shots of the single crystal during and after growth by Czochralski method. (**E**) Laue diffraction pattern emphasizing the single crystalline nature of the as grown MoSi$_2$ single crystal.

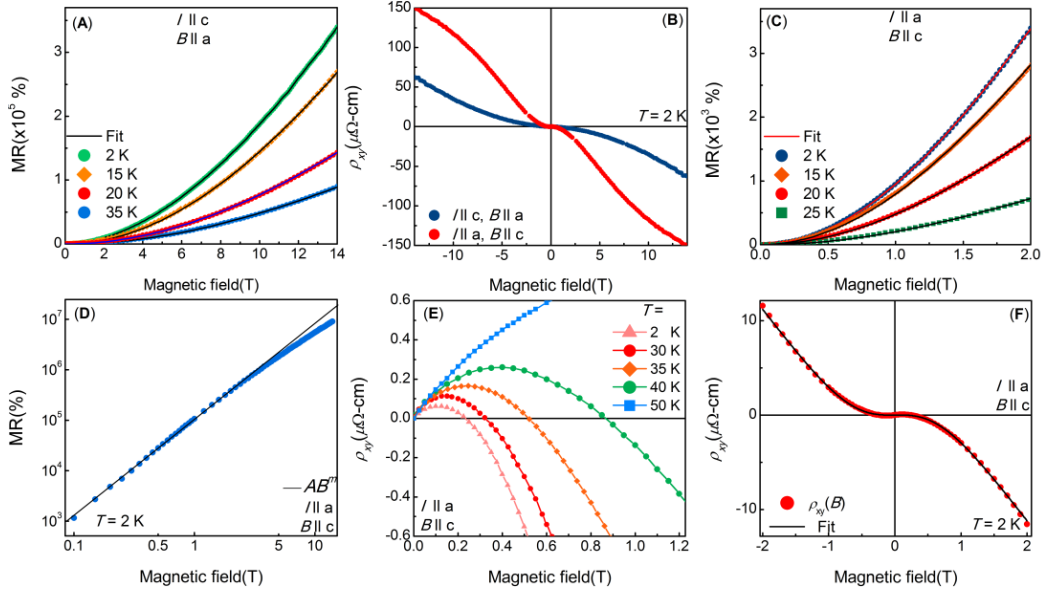

**Fig. 4. Magnetotransport properties of MoSi$_2$ and the validation of the semiclassical theory for two-band transport**. (**A**) Fitting of the transverse MR for $I \parallel c$ based on two-band model eqn. (1) at few representative temperatures. The solid lines represent the theoretical fits. (**B**) The field dependence of Hall resistivity $\rho_{xy}$ in at 2 K. (**C**) Fitting of the transverse MR for $I \parallel a$ based on eqn. (1) at few representative temperatures. (**D**) The field dependence of MR at 2 K for $I \parallel a$, plotted in *log-log* scales to demonstrate the mild saturation tendency in high-field regime above 5 T. (**E**) $\rho_{xy}(B)$ for $I \parallel a$ and $B \parallel c$ at few selected temperatures, showing the sign reversal phenomenon. (**F**) Fitting of the 2 K-$\rho_{xy}(B)$ for $I \parallel a$ and $B \parallel c$ based on the two-band model eqn. (2).

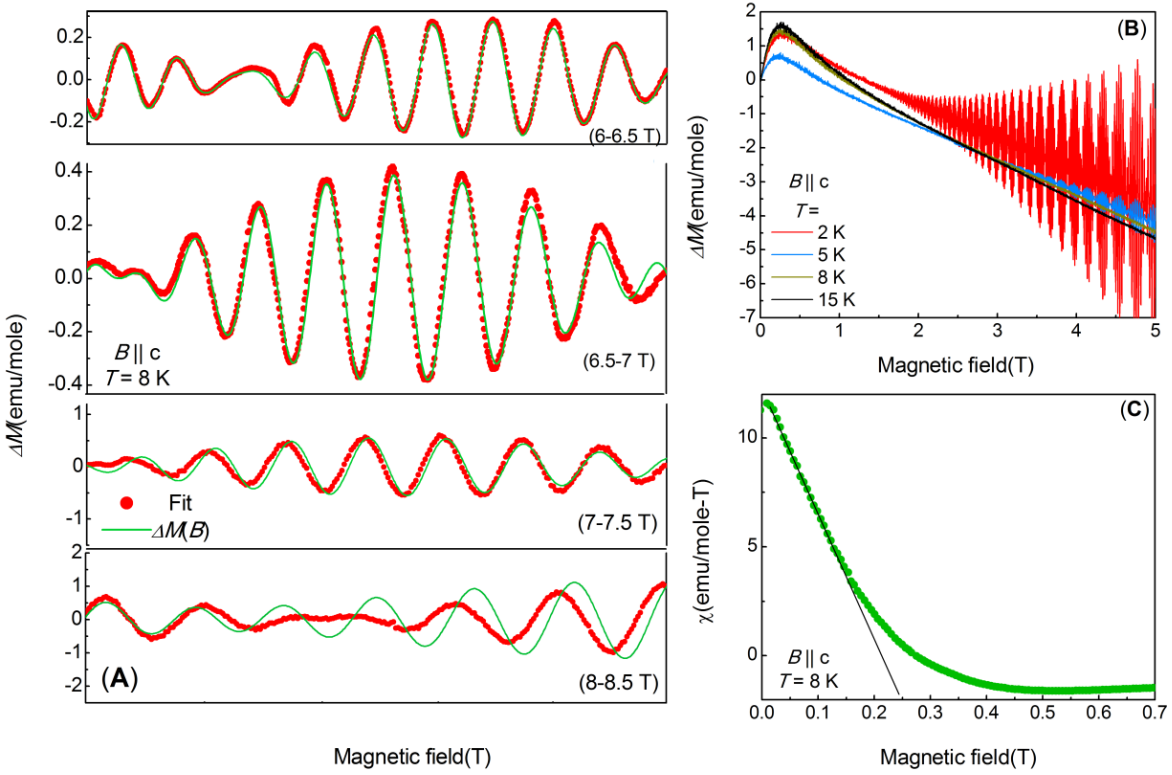

**Fig. 5. The dHvA oscillations and low-field paramagnetic singularity of magnetization in MoSi$_2$**. (**A**) The dHvA oscillations of magnetization at 8 K (solid symbols). The magnetic field ranges are indicated in the braces. Green solid line represents the theoretical fit based on the multiple Lifshitz-Kosevich function (eqn. 4). (**B**) Field dependence of magnetization of MoSi$_2$ at few representative temperatures between 2 and 15 K, showing the paramagnetic peak near zero field. (**C**) The differential susceptibility $\chi$ ($=dM/dB$) as a function of magnetic field, demonstrating the paramagnetic singularity near zero field. The dash-line represents a linear fit to the experimental data.

**Table 1. Estimated parameters from dHvA oscillations.**

| Frequency (T) | $m^*/m_e$ | $T_D$ (K) | $\tau_Q$ (s) | $\mu_Q$ (cm$^2$/V-s) |
|---|---|---|---|---|
| 725 | 0.29 | 3.2 | $3.8\times10^{-13}$ | $2.30\times10^3$ |
| 825 | 0.32 | 2.8 | $4.2\times10^{-13}$ | $2.33\times10^3$ |

# Supplementary Materials

**Fig. S1: Magnetic field dependence of the transverse magnetoresistance for $I\|c$.**

Fig. S1 shows the magnetic field dependence of the transverse MR measured as at several constant temperatures between 2 and 100 K with current along the *c*-axis. The transverse MR for $I\|c$ responds to magnetic field in a way very similar to that for $I\|a,$ at least in qualitative level. The non-saturating behavior of the MR persists up to the highest applied magnetic field (here 14 T). The MR decreases slowly with increasing temperature from 2 to 15 K, but drops quickly at higher temperatures. The MR at 100 K reduces drastically to a value of only 646 % in 14 T magnetic field.

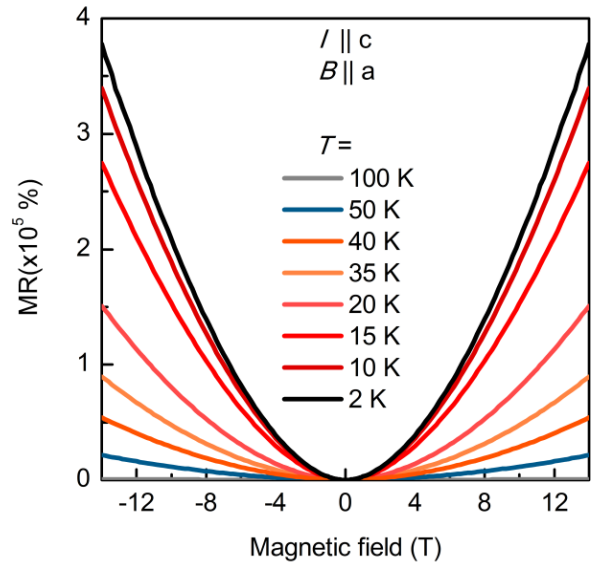

**Fig. S1.** Magnetic field dependence of the transverse magnetoresistance for $I\|c$ at a few representative temperatures between 2 and 100 K.

**Fig. S2: Subhnikov-de Haas (SdH) quantum oscillations in MoSi$_2$ at 2 K.**

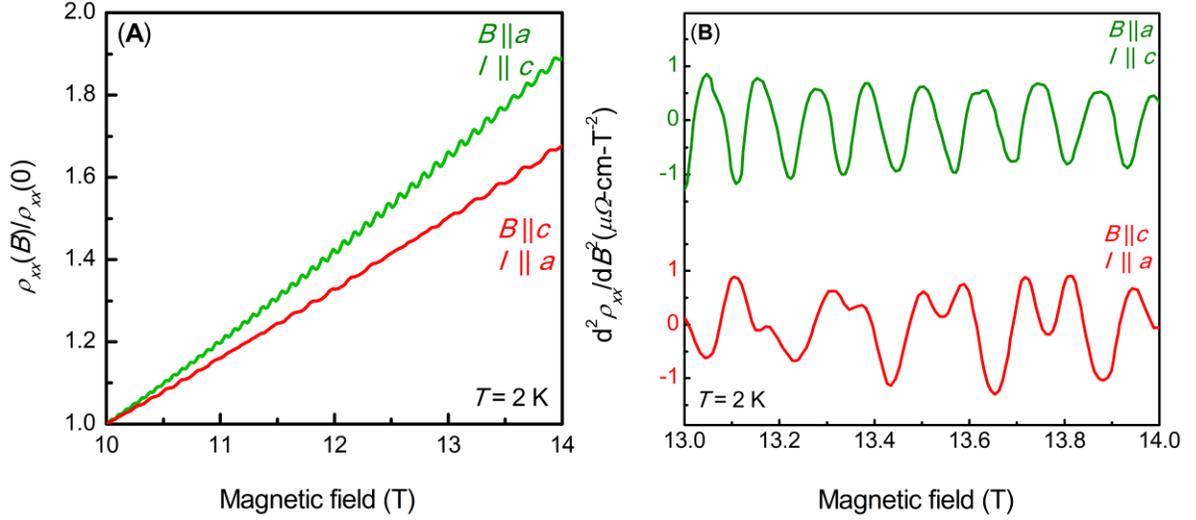

**Fig. S2.** (**A**) $\rho_{xx}$ (B) data for MoSi$_2$ at 2 K, exhibiting the SdH oscillations. (**B**) Purely oscillatory component of the $\rho_{xx}$ (B) data at 2 K, obtained by performing a second-order derivative of $\rho_{xx}$ with respect to the magnetic field.

As shown in the panel (A) of Fig. S2 the magnetic field dependence of the electrical resistivity $\rho_{xx}$ measured in transverse field geometry, exhibits Subhnikov-de Haas (SdH) quantum oscillations in high-field regime. The oscillatory component is separated from the background resistivity by performing a second-order derivative of $\rho_{xx}$ with respect to the magnetic field. The obtained SdH oscillations in high-field regime are shown in the panel (B) of Fig. S2. For $B||a$, the SdH oscillations exhibit conventional beating pattern without any notion of peak splitting and/or broadening (green solid line), indicating that the Zeeman effect is rather trivial when the magnetic field is aligned along the *a*-axis (green solid line). On the other hand, strong Zeeman effect manifests itself as splitting of SdH oscillation peaks for $B||c$ (red solid line), consistent with the dHvA results presented in Fig. 2B.